\documentclass[11pt]{article}
\usepackage{times}
\usepackage{geometry}
\usepackage{floatrow}
\usepackage{float}
\usepackage{hyperref}
\usepackage{graphics,graphicx}
\usepackage{subcaption}
\usepackage{epsf,wrapfig}
\usepackage{caption}
\usepackage[rightcaption]{sidecap}
\usepackage{amsmath}

\usepackage{natbib,bm}

\usepackage{paralist}
\let\olditem\item
\renewenvironment{thebibliography}[1]{%
  \section*{\refname}
  \let\par\relax
  \renewcommand{\item}[1][]{\olditem[\textbullet]}%
  \inparaenum}{\endinparaenum}

\newcounter{aff}

\DeclareGraphicsExtensions{.jpg,.pdf,.png,.eps}

\usepackage{xcolor}
\usepackage{dirtytalk}
\usepackage{tcolorbox}

\bibpunct{(}{)}{;}{a}{}{,}

\usepackage{soul}

\definecolor{darkcyan}{rgb}{0.0, 0.0, 0.7}

\definecolor{boxcolor}{RGB}{193, 229, 255}
\definecolor{figboxcolor}{RGB}{193, 229, 255}
\definecolor{boxcolorgray}{RGB}{227, 227, 227}
\definecolor{darkcyan}{rgb}{0.0, 0.0, 0.7}

\sethlcolor{boxcolorgray}

\def\gtrsim{\lower 2pt \hbox{$\, \buildrel {\scriptstyle >}\over
{\scriptstyle \sim}\,$}}
\def\lesssim{\lower 2pt \hbox{$\, \buildrel {\scriptstyle <}\over
{\scriptstyle \sim}\,$}}

\def\approxlt{\lower.2em\hbox{$\buildrel < \over \sim$}}
\def\approxgt{\lower.2em\hbox{$\buildrel > \over \sim$}}

\def\be{\begin{equation}}
\def\ee{\end{equation}}
\def\bea{\begin{eqnarray*}}
\def\eea{\end{eqnarray*}}

\def\jref@jnl#1{{\rm#1}}
\def\aj{\jref@jnl{AJ}}			
\def\araa{\jref@jnl{ARA\&A}}		
\def\apj{\jref@jnl{ApJ}}			
\def\apjl{\jref@jnl{ApJ}}		
\def\apjs{\jref@jnl{ApJS}}		
\def\ao{\jref@jnl{Appl.~Opt.}}		
\def\apss{\jref@jnl{Ap\&SS}}		
\def\aap{\jref@jnl{A\&A}}		
\def\aapr{\jref@jnl{A\&A~Rev.}}		
\def\aaps{\jref@jnl{A\&AS}}		
\def\azh{\jref@jnl{AZh}}			
\def\baas{\jref@jnl{BAAS}}		
\def\jrasc{\jref@jnl{JRASC}}		
\def\memras{\jref@jnl{MmRAS}}		
\def\mnras{\jref@jnl{MNRAS}}		
\def\pra{\jref@jnl{Phys.~Rev.~A}}	
\def\prb{\jref@jnl{Phys.~Rev.~B}}	
\def\prc{\jref@jnl{Phys.~Rev.~C}}	
\def\prd{\jref@jnl{Phys.~Rev.~D}}	
\def\pre{\jref@jnl{Phys.~Rev.~E}}	
\def\prl{\jref@jnl{Phys.~Rev.~Lett.}}	
\def\pasp{\jref@jnl{PASP}}		
\def\pasj{\jref@jnl{PASJ}}		
\def\qjras{\jref@jnl{QJRAS}}		
\def\skytel{\jref@jnl{S\&T}}		
\def\solphys{\jref@jnl{Sol.~Phys.}}	
\def\sovast{\jref@jnl{Soviet~Ast.}}	
\def\ssr{\jref@jnl{Space~Sci.~Rev.}}	
\def\zap{\jref@jnl{ZAp}}			
\def\nat{\jref@jnl{Nature}}		
\def\iaucirc{\jref@jnl{IAU~Circ.}}
\def\aplett{\jref@jnl{Astrophys.~Lett.}}
\def\apspr{\jref@jnl{Astrophys.~Space~Phys.~Res.}}
\def\bain{\jref@jnl{Bull.~Astron.~Inst.~Netherlands}}
\def\fcp{\jref@jnl{Fund.~Cosmic~Phys.}}
\def\gca{\jref@jnl{Geochim.~Cosmochim.~Acta}}
\def\grl{\jref@jnl{Geophys.~Res.~Lett.}}
\def\jcp{\jref@jnl{J.~Chem.~Phys.}}	
\def\jgr{\jref@jnl{J.~Geophys.~Res.}}	
\def\jqsrt{\jref@jnl{J.~Quant.~Spec.~Radiat.~Transf.}}
\def\memsai{\jref@jnl{Mem.~Soc.~Astron.~Italiana}}
\def\nphysa{\jref@jnl{Nucl.~Phys.~A}}
\def\physrep{\jref@jnl{Phys.~Rep.}}
\def\physscr{\jref@jnl{Phys.~Scr}}
\def\planss{\jref@jnl{Planet.~Space~Sci.}}
\def\procspie{\jref@jnl{Proc.~SPIE}}

\def\ion#1#2{#1$\;${\small\rm\@Roman{#2}}\relax}

\newcount\lecurrentfam
\def\LaTeX{\lecurrentfam=\the\fam \leavevmode L\raise.42ex
\hbox{$\fam\lecurrentfam\scriptstyle\kern-.3em A$}\kern-.15em\TeX}
\def\sizrpt{
(\fontname\the\font): em=\the\fontdimen6\font, ex=\the\fontdimen5\font
\typeout{
(\fontname\the\font): em=\the\fontdimen6\font, ex=\the\fontdimen5\font
}}
\definecolor{dark}{rgb}{ 0.34, 0.09, 0.49}

\newtcolorbox{purplebox}{
    colback=purple!8,
    colframe=purple!8,
    boxrule=0pt,
    sharp corners,
    left=6pt,
    right=6pt,
    top=6pt,
    bottom=6pt
}

\usepackage{fancyhdr,url}
\usepackage{transparent}
\usepackage{tablefootnote}
\usepackage{mathpazo}

\begin{document}
\vspace{-3em}

\noindent
\begin{raggedright}
{\Huge\bfseries From Scaling Relations to Physical Laws: \\
\Large Constraining the Physics of Star Formation Across Local Galaxies 
}  \end{raggedright}\\

\noindent
{
Jakob den Brok$^1$ , Cecilia Bacchini$^{2,3}$, Leindert A. Boogaard$^{4}$, Mathilde Bouvier$^{5,4}$, Viviana Casasola$^{6}$, Laura Colzi$^{7}$, Timothy A. Davis$^8$, Cosima Eibensteiner$^{9}$, Daniel Espada$^{10,11}$, Simon Glover$^{12}$, Logan H.\ Jones$^{13}$, Ryan Keenan$^{1}$, Ralf S.\ Klessen$^{12,14}$, Ivanna\ Langan$^{15}$, Dieu D.\ Nguyen$^{16}$, Miguel Pereira-Santaella$^{17}$, Miguel Querejeta$^{18}$, Ilaria Ruffa$^{6, 19}$, Francesco Salvestrini$^{20,21}$, Jiayi Sun$^{22}$, Sabine Thater$^{23}$, Robin Tress$^{1}$, Antonio Usero$^{18}$, Thomas G. Williams$^{24}$
}

\vspace{1em}
\noindent {\footnotesize{\textit{(Affiliations can be found after the references)}}}

\begin{tcolorbox}[boxsep=3pt,left=0pt,right=0pt,top=1pt,bottom=1pt,colframe=boxcolor,colback=boxcolor]

\justify
\textbf{Abstract:} 
The conversion of cold dense gas, traced by molecular line emission,  into stars remains one of the fundamental unsolved problems in galaxy evolution. Observations over the past decades have established empirical relations between gas and star formation, yet their physical origin remains uncertain. Gravity, turbulence, and stellar feedback as main drivers provide fundamentally different pictures of how star formation is regulated, while reproducing remarkably similar galaxy-scale scaling relations. To disentangle the individual importance of each process, not only its mass and distribution, but also its temperature, density, excitation, and chemistry must be probed. However, current or planned (sub)millimetre facilities cannot combine statistical breadth, cloud-scale resolution, and multi-line diagnostics across large galaxy samples. A new, flagship interferometer with increased line sensitivity and bandwidth enables obtaining multi-transition, multi-isotopologue observations of molecular gas at GMC-resolved scales across nearby galaxies, enabling millions of independent measurements of its physical state across a wide range of extragalactic environments. This will transform molecular gas from an empirical tracer into a quantitative probe of the physical processes governing the relation between the dense gas and star formation. With the 2040s landscape of optical and radio facilities, ALMA2040 will facilitate a predictive theory of the conversion of gas into stars.

\end{tcolorbox}

\newpage{}

\vspace{-0.7cm}
\section{Scientific context and open questions}
\vspace{-0.2cm}

A predictive theory of how cold dense gas converts into stars remains one of the central problems in astrophysics. Although observations over the past decades have established a series of empirical scaling relations connecting molecular gas, galaxies, and star formation, the underlying physical processes that give rise to these relations remain uncertain. These include the Kennicutt–Schmidt relation  \citep{Kennicutt1998,Kennicutt2012}, Larson-type cloud scaling relations \citep{Larson1981,Heyer2009}, and the molecular gas main sequence linking molecular gas and stellar mass. At the same time, the physical state of the gas depends on cooling processes ranging from molecular and atomic line emission at lower densities to gas--dust coupling at high densities \citep{Klessen2016}, while the molecular gas itself is not necessarily fully traced by CO because of CO-dark H$_2$ and the uncertain CO-to-H$_2$ conversion factor, $\alpha_{\rm CO}$. These empirical relations are still widely used to calibrate sub-grid star-formation prescriptions in galaxy simulations \citep{Vogelsberger2014,Romano2017, Feldmann2026}, while different physical mechanisms, including gravitational collapse, turbulence, and stellar feedback, can reproduce the observed relations despite invoking fundamentally different regulation mechanisms.\vspace{-2mm}

\begin{tcolorbox}[boxsep=3pt,left=0pt,right=0pt,top=1pt,bottom=1pt,colframe=boxcolor,colback=boxcolor]
A critical next step is to systematically identify the physical state of the gas from which these scaling relations emerge, rather than treating them as universal laws. 
Multi-line molecular spectroscopy provides one of the most direct and powerful means of constraining the physical state of the molecular ISM (including its temperature, density, optical depth, chemistry, and dynamical state) and thereby identifying the physical parameters that govern star formation.
\end{tcolorbox}

\vspace{-7mm}

\subsection{Different theories of star formation}
\vspace{-0.8em}

The wealth of multi-wavelength observations accumulated over the past decades, together with increasingly sophisticated numerical simulations, has led to different physically-motivated theories for the key drivers that regulate how molecular gas is converted into stars \citep{McKee2007, Krumholz2012, Padoan2014, Klessen2016}. Although these theories differ in the physical processes assumed to regulate star formation, the scaling relation emerges schematically as expressed by the following equation relating the SFR surface density, $\Sigma_{\rm SFR}$ (mass per unit area per unit time), to the gas surface density, $\Sigma_{\rm gas}$ (mass per unit area), via the ratio of the star-forming efficiency, $\eta_{\rm SF}$, i.e. the fraction of the molecular gas that is converted into stars, and the characteristic timescale, $\tau$, of the regulating process:
\vspace{-0.8em}

\begin{equation}\label{eq:simple}
    \Sigma_{\rm SFR} \propto \frac{\eta_{\rm SF}}{\tau}\cdot \Sigma_{\rm gas}.
\end{equation}
\vspace{-1.2em}

Gravity, turbulence, stellar feedback, and cooling processes all influence and impact to a different degree the evolution of molecular clouds: Collapse-dominated models associate star formation directly with the amount of gas undergoing gravitational collapse over a free-fall time set by the molecular gas density \citep[e.g.,][]{Elmegreen2002}. Turbulence-regulated models propose that supersonic turbulence determines the fraction of gas that exceeds a critical density for collapse \citep{MacLow2004, Krumholz2005, Padoan2011, Federrath2012}. Feedback-regulated models instead argue that star formation self-adjusts until stellar feedback balances the gravitational weight of the ISM \citep{Ostriker2011, Ostriker2022}. These physically distinct theories all reproduce approximately the observed Kennicutt–Schmidt relation. 
On galaxy scales (i.e. ${>}\rm kpc$), averaging over many clouds and over timescales comparable to the cloud lifecycle suppresses much of the intrinsic variation in the underlying physical state. Moreover, quantities such as density,  turbulent support, and feedback strength correlate with gas surface density and galaxy dynamical properties. Consequently, physically different models can project onto similar relations between $\Sigma_{\rm gas}$ and $\Sigma_{\rm SFR}$, even though they predict different multidimensional distributions of the underlying gas properties.\medskip

The difference of these theories lies not in their ability to reproduce galaxy-scale scaling relations, but in their predictions for the physical state of the molecular gas \citep{Hopkins2013}. Gravity, turbulence, and stellar feedback are not mutually exclusive ingredients of star formation. Rather, competing theoretical frameworks differ in which process primarily sets the fraction of gas participating in collapse and the timescale over which it is converted into stars. 
They make fundamentally different predictions for the internal structure and dynamical state of giant molecular clouds. For example, whether clouds are approximately virialised and turbulence-supported, with only a small fraction of gas reaching gravitationally unstable densities, or instead undergo global hierarchical collapse with an increasing dense-gas fraction \citep[e.g.][]{Krumholz2005,Padoan2011,Federrath2012}, and thus for the star-forming efficiency and timescale in Equation \ref{eq:simple}.
\textbf{Measuring these quantities simultaneously across statistically representative galaxy samples at GMC-scales (${\le}100$pc) therefore provides one of the few direct observational routes for discriminating between competing theories of star formation}.\medskip

Indeed, we expect the physical conditions of the ISM to vary. For instance, the molecular gas kinetic temperatures range from ${\sim}10$\,K in embedded molecular clouds but can also reach ${>}100\,$K in very active star-forming environments, such as hot cores forming stars. Moreover, the average H$_2$ volume density on $10-{100}\,$pc scales is reported to vary by orders of magnitude from $10^1-10^5$\,cm$^{-3}$ \citep{Tanaka2024}.  This range in underlying physical conditions not only affects (or drives) the origin of the empirical scaling relations, but also directly affects the resulting stellar populations that are being formed. 
Evidence of this is that the cluster formation efficiency (i.e. the fraction of stars that form in clusters) has been reported to vary as a function of environment \citep{Adamo2015,Ginsburg2018}. Also, the initial mass function itself is believed to change in more massive early type galaxies or starbursts \citep{Cappellari2016,Brown2019}. 
\vspace{-10mm}




\subsection{Constraining the key physics of star formation through multi-line observations}
\vspace{-2mm}

CO emission provides a powerful tracer of molecular gas mass and kinematics, and in particular the CO(1-0) and CO(2-1) transitions have become the workhorse tracers for extragalactic molecular ISM observations \citep{Bolatto2013}. But the observed intensity of an individual transition depends simultaneously, so degenerately, on gas density, temperature, and opacity \citep{Klessen2016}. Thus the physical variables that distinguish competing theoretical star-formation drivers are also largely inaccessible through observations of a single molecular transition. On the contrary, they can be accessed with multi-transition observations because the various molecular lines are sensitive to different degrees of the underlying physical state of the gas.\medskip

Multi-transition and multi-isotopologue molecular spectroscopy enable us to break these degeneracies as different rotational transitions probe different excitation conditions, while isotopologues with different abundances and optical depths constrain opacity and column density. Together, observations of  $^{12}$CO,  $^{13}$CO, C$^{18}$O and complementary dense-gas tracers (such as HCN, HCO$^+$ and others; see simulations by \citealt{Priestley2024} of star-forming clouds) can constrain the temperature and density structure of molecular gas by solving numerically or analytically (under the assumption of LTE) the radiative transfer equations. For example, a commonly used non-LTE radiative transfer code is \texttt{RADEX} \citep{vanderTak2007}, that models molecular line intensities as a function of the physical state of the molecular gas \citep[e.g.,][]{Teng2022, Teng2023, Zhang2025} and can be matched with observations. Similarly, the \texttt{UCLCHEM} software tool  \citep{Holdship2017,Vermarien2026}, further assesses the relative chemical abundances and  the chemical evolution in the presence of shocks, as in the case of star forming regions. The resulting physical measurements can then be combined with velocity information to determine quantities such as the virial parameter, free-fall time, turbulent Mach number, and molecular-gas conversion factor. However, current large ($N_{\rm gal}{>}50$) galaxy surveys are generally limited by the small (generally 1--2) number of molecular lines, making accurate and systematic constraints challenging.

\vspace{-0.6em}
\begin{tcolorbox}[boxsep=1pt,left=0pt,right=0pt,top=1pt,bottom=1pt,colframe=boxcolor,colback=boxcolor]
Multi-line mapping will enable us to quantitatively link density, excitation, and chemical abundances to galaxy structure and evolution. Inferring the density distribution requires observing lines that span a wide range of effective critical densities and excitation temperatures. To constrain excitation, one observes multiple transitions of the same species. Realistically, CO isotopologue lines and dense gas tracers, such as HCN and HCO$^{+}$ are ${\sim}10{-}50$x fainter than the bright $^{12}$CO lines. Therefore, the significant leap in line sensitivity by ALMA2040 will provide access to multi-line observations, with a wide range of densities and temperatures required to constrain the physical state of the gas through non-LTE radiative transfer modeling.

\end{tcolorbox}
\vspace{-2em}

\subsection{From individual galaxies to a statistical theory of the molecular ISM}
\vspace{-0.5em}

The field is now entering a new regime in which key aspects of the molecular ISM can be studied statistically rather than through isolated case studies, such as the bulk distribution and dynamic state: large surveys have established the molecular-gas content of representative galaxy populations at kpc-scale (EDGE, \citealt{Bolatto2017}; COMING, \citealt{Sorai2019}, ALMAQUEST, \citealt{Lin2020}), while recent programs such as PHANGS \citep{Leroy2021} or WISDOM \citep{Davis2022} are now resolving molecular structures on ${\sim}$100 pc (or smaller) scales across nearby galaxies. More recently, surveys such as KILOGAS \citep[][in prep.]{Saintonge2026} are extending molecular-gas mapping at kpc-scale beyond massive-star-forming galaxies, providing an unprecedented view of how molecular-gas reservoirs vary across galaxy mass, morphology, star-formation, nuclear activity, and environment. In particular extending to a wider range of galaxy types will provide exploring the impact of a much more diverse range of extragalactic environments.
Existing GMC-scale (${\le100}\,$pc) multi-line studies remain rare and necessarily focus on a limited number of nearby and often exceptionally bright galaxies. The surveys that provide GMC-scale multi-CO imaging generally lack the sensitivity required to obtain a comprehensive molecular-line inventory. The result is a fundamental mismatch between the scale on which theories make predictions and the scale on which observations can currently test them.
\medskip 

Current multi-line surveys reveal that the molecular gas properties vary systematically with galactic environment, including stellar surface density, dynamics, metallicity, pressure, and proximity to galactic centres \citep{Teng2022,denBrok2025}. Inferring molecular gas surface densities, however, requires conversion factors that themselves depend on the physical and chemical conditions of the gas. The same inferred molecular gas surface density can therefore correspond to substantially different physical conditions. Multi-transition observations provide the means to constrain these conditions, improving molecular gas mass estimates more generally. Ultimately, ALMA2040 could reach the stage where true H$_2$ abundances can be calculated through high-resolution virial dynamics assumptions or chemically calibrated. 
\medskip


A predictive theory of star formation and the physical origin of the scaling relations requires moving from individual galaxies to a statistically-representative population while retaining the physical resolution required to resolve molecular gas structures. Measuring the molecular-gas temperature, density, excitation, chemistry (see also ALMA2040-WP by Bouvier et al.), turbulence, and gravitational state for thousands of GMCs spanning the full diversity of galaxy environments is key. Such a dataset would enable an empirical determination of how physical conditions vary across clouds and environments, and whether the physical laws governing star formation are universal or vary systematically across galaxies. Current facilities cannot simultaneously provide this combination of sample size, GMC-scale resolution, sensitivity, and multi-line spectral coverage. \textbf{ALMA2040 would transform the study of star formation from studies of individual galaxies and clouds into a statistical experiment that retains cloud-scale physics while testing the universality and origin of the laws governing the conversion of molecular gas into stars}.


\begin{tcolorbox}[boxsep=0pt,left=0pt,right=0pt,top=0pt,bottom=0pt,colframe=boxcolor,colback=boxcolor]
\subsection{Key Science Questions to adress in the 2040s} \vspace{-0.5em}
We propose the following science questions for the next flagship (sub)mm interferometer:    \vspace{-0.5em}
\begin{enumerate}
    \item What physical state does molecular gas occupy across the galaxy population?
    \vspace{-0.5em}
    \item Which physical processes drive and regulate the conversion of molecular gas into stars?    \vspace{-0.5em}
    \item How does this regulation depend on galaxy environment?    \vspace{-0.5em}
\end{enumerate}
In combination with new observations from current and upcoming facilities with measurements that trace recent star formation and stellar feedback, the derivation of the physical state of the molecular gas will result in an observationally-vetted general theory of star formation in galaxies which is fundamental for modeling galaxy formation and evolution.
\end{tcolorbox}
\vspace{-2em}

\section{ALMA2040: Technical Requirements}\label{sec:requirements}
\vspace{-0.3cm}

Observations of multiple molecular lines are powerful tools to investigate the physical conditions of the gas in various environments. Until now, as outlined above, systematic multi-CO line studies towards nearby galaxies have been mostly limited to galactic centers \citep{Israel2020, Teng2023}, extreme starbursts \citep[e.g.,][]{Sliwa2017}, or active galactic centers \citep{Oosterloo2017, Ruffa2022},  where mm lines are brighter and thus more easily observed. Outside these environments, observations remain extremely expensive (${>}100\,$h for a single galaxy).  
\textbf{Critically, systematic, GMC-resolved (${<}50$\,pc) multi-line surveys will remain prohibitively expensive for the broader nearby galaxy population, leaving a major gap in our ability to generalize findings beyond a handful of well-studied systems.} In essence, the major limitation for this
science remains the inefficiency of current mm observations to map entire galaxies, due to: (i) low sensitivity at
long baselines, (ii) small field-of-view, and (iii) lack of multi-band (i.e obtaining the spectrum across multiple \emph{atmospheric windows}) capabilities, necessitating multiple, time-intensive observations. Overcoming these limitations with ALMA2040 will be critical to enabling a systematic,
unbiased census of GMC-resolved molecular gas properties across the local universe.
\vspace{-0.4cm}

\subsection{Probing the physics and resolving the star-forming reservoir at ${<}100\,$pc}
\vspace{-0.1cm}

The current state of the art systematic molecular-gas surveys of nearby galaxies reach ${\sim}100$\,pc resolution. The PHANGS--ALMA survey, for example, mapped CO(2--1) in 90 nearby star-forming galaxies at approximately 1$''$ resolution, corresponding to ${\sim}100$\,pc \citep{Leroy2021}. At this resolution, individual resolution elements approach the characteristic size of giant molecular clouds (GMCs), enabling the statistical characterization of their masses, sizes, linewidths, surface densities, and demographics \citep{Sun2020,Leroy2025}. This represents a major advance over kpc-scale observations, which average over many molecular clouds and star-forming regions.\medskip

\begin{figure}[h]
    \centering
    \includegraphics[width=\linewidth]{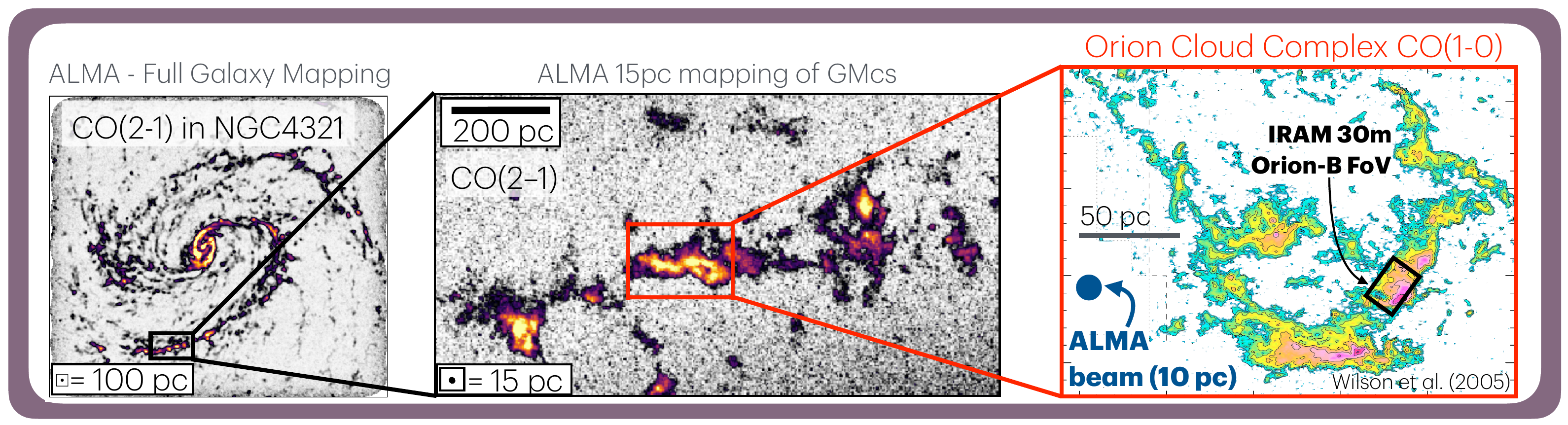}
    \vspace{-.3cm}
    \caption{ALMA already reaches the required resolution (central panel), but only for small selected regions and only in CO emission. The transformational requirement is to combine this resolution with multi-line sensitivity and galaxy-wide mapping speed. Mapping the molecular gas at ${\sim}10\,$pc resolution enables direct comparison to MW cloud studies. For reference, the right panel shows the Orion B cloud complex, which at 10\,pc resolution would be sampled in multiple beams. Such extragalactic resolved observations enable placing GMCs correctly in the environmental context. \vspace{-0.4cm}}
    \label{fig:highres}
\end{figure}

However, ${\sim}100$\,pc should be regarded as a \emph{cloud-scale} rather than a true cloud-resolving regime. The ISM is structured at sub-100 pc scales: it contains filaments, dense clumps, cores, and embedded star-forming regions spanning scales of a few parsecs to tens of parsecs. At ${\sim}100$\,pc resolution, these structures are blended together, and the measured temperature, density, linewidth, and star-formation efficiency represent averages over multiple physical stages of the cloud lifecycle. Indeed, PHANGS observations have shown that molecular clouds and young stellar regions are often spatially offset on these scales, revealing an evolutionary sequence that would be obscured at kpc resolution \citep{Kruijssen2019,Chevance2020}. Resolving scales of $\sim5{-}20$\,pc provides qualitatively different access to the star-forming reservoir (see WP by Duarte Cabral et al.), approaching the sizes of individual cloud substructures and H\textsc{II} regions and beginning to separate the dense, actively star-forming gas from the more extended molecular reservoir. It would also enable a direct connection to molecular gas studies in the Milky Way, as these scales approach the size of typical clouds, such as Taurus or the Orion B cloud complex sub-features (see \autoref{fig:highres}).\medskip 

Despite the scientific importance of these scales, systematic observations of entire galaxies at ${\sim}5{-}20$\,pc resolution remain prohibitively expensive. The combination of angular resolution, surface-brightness sensitivity, and mapping speed required to recover faint molecular emission over the full extent of a nearby galaxy is beyond the capabilities of current interferometric surveys. Consequently, sub-$100$\,pc studies have so far largely been restricted to selected regions of individual galaxies \citep[e.g., ALCHEMI][]{Martin2021}. No existing survey simultaneously resolves molecular-cloud substructure and samples a representative population of external galaxies.

\hrule
\hrule
\vspace{-1.2em}

\subsection{High sensitivity to map fainter molecular gas tracers beyond $^{12}$CO} 
\vspace{-0.8em}

Robustly constraining the physical state of molecular gas requires the simultaneous detection of multiple molecular transitions and isotopologues of CO. The bright low-$J$ transitions of $^{12}$CO are the most practical tracers for high-resolution mapping across entire galaxies, but the diagnostics provided by rarer isotopologues are substantially fainter. In particular, $^{13}$CO is typically ${\sim}10{-}20$ times fainter than $^{12}$CO in brightness temperature, while C$^{18}$O and higher-excitation transitions can be considerably fainter still \citep{Wilson1999,Jimenez2019,Leroy2021}. These faint lines nevertheless provide crucial constraints on column density, optical depth, excitation, temperature, and density that cannot be obtained from $^{12}$CO alone.\medskip

CO and its isotopologues probe the bulk molecular gas distribution, while higher–dipole moment species such as HCN, HNC, and HCO$^+$ trace denser gas more closely associated with star-forming cloud cores \citep{Gao2004}. This is particularly important in high-pressure environments, such as galaxy centres and starbursts, where much of the gas may already exceed the effective critical density of CO, making CO alone insufficient to identify the star-forming component. More selective tracers such as N$_2$H$^+$ can instead probe cold, dense gas associated with gravitationally unstable cores \citep{Bergin2007} and potentially connect observations more directly to a volume-density threshold for star formation \citep{Priestley2024}. Crucially, these dense gas tracers provide a key complementary diagnostic to benchmark the physical state measurements derived from CO isotopologue lines across a range of extragalactic environments.
\vspace{-1em}

\subsection{Obtaining high-fidelity, total flux recovered mapping observations}\vspace{-3mm}
Observations of nearby galaxies show that a substantial fraction of molecular line emission is distributed on angular scales that are resolved out by interferometers. For example, \citet{Pety2013} found that ${\sim}50\%$ of the CO(1--0) flux of M51 was missing from the PdBI observations before combining them with IRAM 30\,m data. Similar comparisons have found $\sim10{-}50\%$ missing flux in other nearby galaxies \citep{Koda2011,Leroy2021_pipe}, including for relatively faint tracers such as $^{13}$CO(1--0) \citep{Stuber2025}. Recovering these spatial scales is therefore essential both for accurate molecular-gas masses and for high-fidelity imaging of the molecular ISM.\medskip 

Zero-spacing information is particularly important for the multi-line science proposed here. Molecular transitions at different frequencies can have different spatial distributions and surface-brightness sensitivities, while an interferometer samples a different range of spatial scales in units of wavelength for each transition. Consequently, interferometric-only observations can recover different fractions of the emission in different lines, introducing spatially dependent biases into line ratios and hence into the inferred temperature, density, excitation, and column density. Combining interferometric and total-power information ensures that the same physical structures are recovered across transitions and that multi-line models are constrained by the complete molecular emission. \textbf{ALMA2040 should therefore provide short-, and zero-spacing information as an integral part of its high-resolution imaging capability, enabling fully sampled, total-flux-recovered molecular-line maps from cloud to galaxy scales. }

\hrule
\begin{SCfigure}[0.7][h]
\caption{Single-pointing of $^{12}$CO and C$^{18}$O(2–1) with  ALMA (top; C6+C3+7m+TP) and ALMA2040 (bottom; H3+H1+TP). The underlying GMC model is based on ALMA CO(2–1) observations of NGC~4321 (2021.1.00728.S; PI: J. Henshaw), rescaled to a distance of 30\,Mpc. The current ALMA observations require 30\,min of total observing time across all configurations, compared to only 2\,min with ALMA2040 .}\label{fig:simulation}
\includegraphics[width=0.65\textwidth]{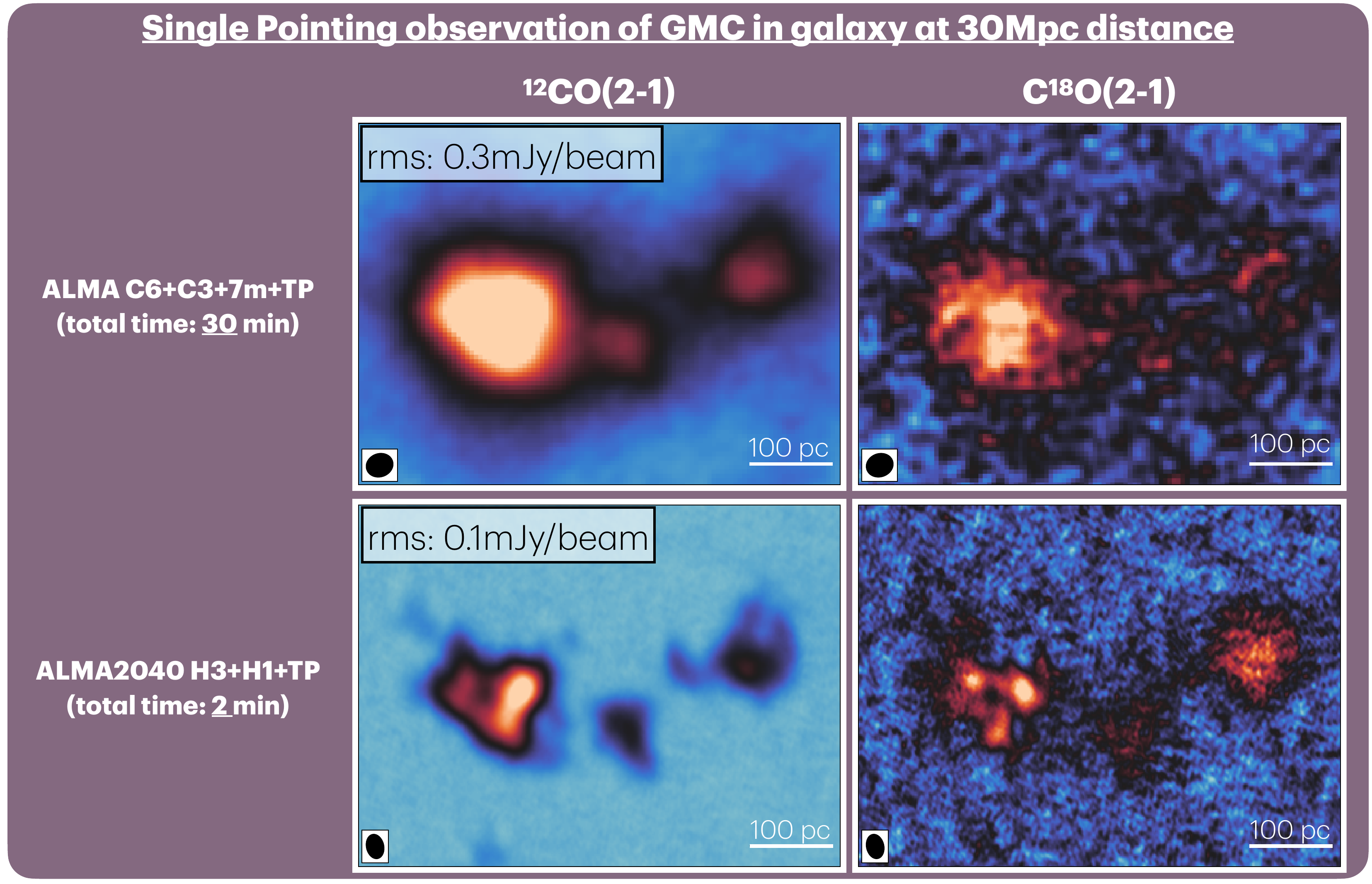}
\end{SCfigure}

\hrule
\subsection{A representative example of ALMA2040 science}
\begin{tcolorbox}[boxsep=0pt,left=0pt,right=0pt,top=0pt,bottom=0pt,colframe=boxcolor,colback=boxcolor]
\textbf{Assumed Capabilities:}
For the ALMA2040 simulations, we assume the following capabilities (the precise capabilities have not yet been finalized, but are expected to be along similar lines):
\vspace{-0.5em}
\begin{itemize}
    \item An array of 255 12-m antennas, configurable into six array configurations (H1--H6) spanning a broad range of angular scales and spatial resolutions.
    \vspace{-0.5em}
    \item Five 25-m single-dish antennas providing zero-spacing information.
    \vspace{-0.5em}
    \item Continuous receiver coverage across the full ALMA frequency range (Bands 1--10).
    \vspace{-0.5em}
    \item A factor of 10 improvement in spectral-line sensitivity relative to the current ALMA 12-m array, enabled by the increased collecting area and improved receiver performance.
\end{itemize}
\end{tcolorbox}

In \autoref{fig:simulation}, we provide a direct comparison of GMC-resolved ALMA and ALMA2040 observations. As input, we use high-resolution (10\,pc) ALMA observations of part of a spiral arm in NGC4321 (2021.1.00728.S; PI: J. Henshaw), rescaled to a distance of 30\,Mpc. The C$^{18}$O(2--1) emission is approximated by dividing the $^{12}$CO map by 30. For both the ALMA and ALMA2040 simulations, we use the ALMA GHOST online tool. Combining ALMA C6+C3+7m+TP (30\,min total) yields a sensitivity of 0.3\,mJy\,beam$^{-1}$ at 100\,km\,s$^{-1}$ and a physical scale of 30\,pc. In contrast, ALMA2040 observations in H3+H1+TP (2\,min total) reaches 0.1\,mJy\,beam$^{-1}$ at 25\,pc.  \medskip

\begin{figure}[h]
    \centering
    \includegraphics[width=0.98\linewidth]{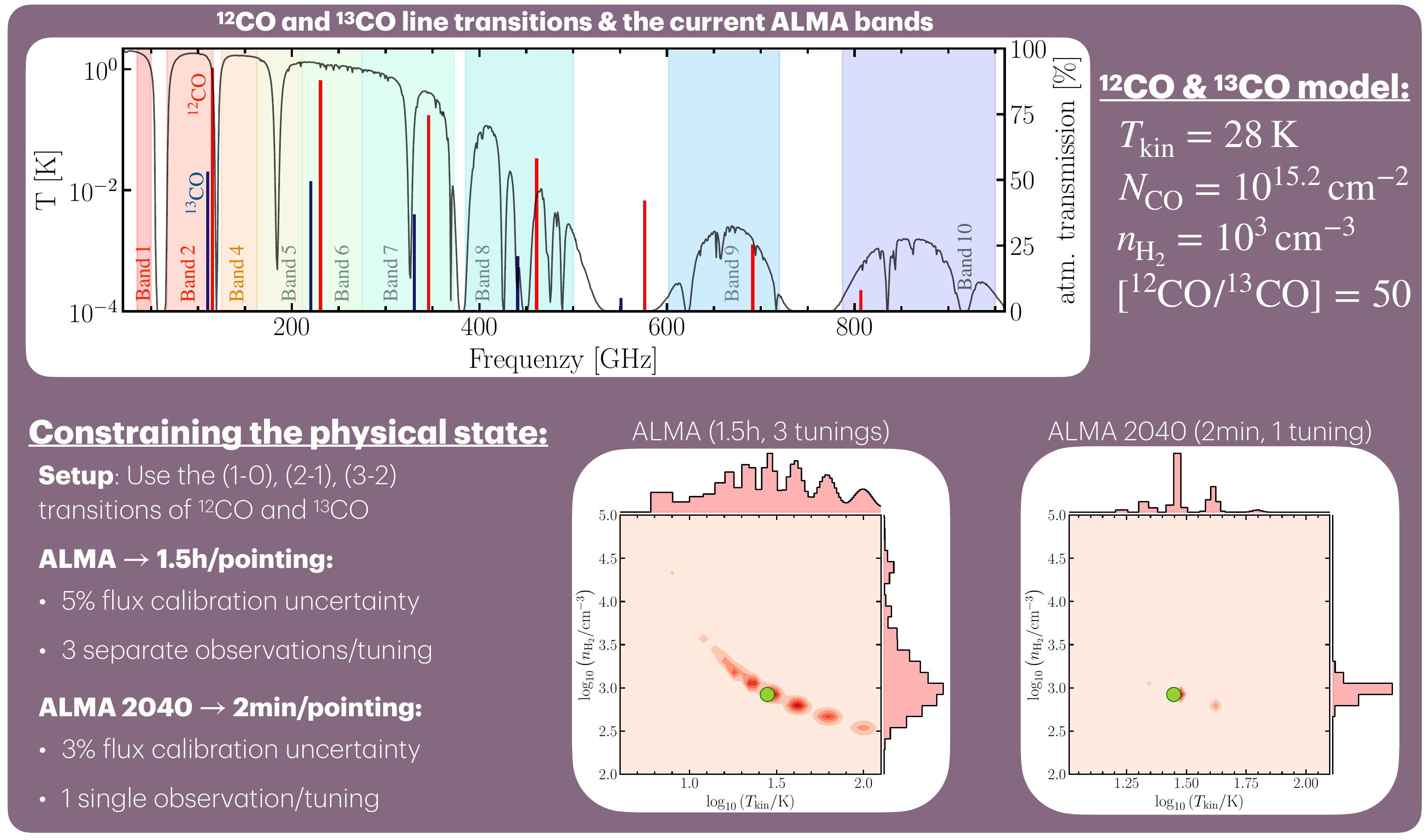}
    \vspace{-.3cm}
    \caption{
    (Top) The $^{12}$CO and $^{13}$CO line brightness across the ALMA frequency range derived using \texttt{RADEX} single-zone modeling under representative physical molecular-gas conditions, illustrating the broad frequency coverage required to constrain the molecular gas at $<100\,$pc resolution. For reference, the atmospheric transmission at 0.5mm is illustrated in blue. (Bottom) Physical constraints from the single-pointing observations as presented in \autoref{fig:simulation}, demonstrating how ALMA2040 has instantaneous constraining power from simultaneous multi-line measurements.
    } 
    \label{fig:simulation2}
\end{figure}

\autoref{fig:simulation2} illustrates the importance of continuous spectral coverage for detecting faint molecular tracers and constraining the physical state of molecular gas. Constraining excitation and chemistry requires multiple transitions and isotopologues spanning a broad frequency range. Using \texttt{RADEX}, we model the expected line emission for representative molecular-gas conditions under a single-zone approximation. We then simulate the corresponding observations, including a flux-calibration uncertainty for each independent tuning. The bottom panels show the resulting marginalized likelihoods for the physical parameters inferred from the simulated ALMA and ALMA2040 observations. Simultaneous observations with ALMA2040 would not only make such multi-line measurements substantially more efficient, but also provide more robust line ratios by avoiding relative calibration uncertainties between separate observing setups. This is particularly important when the intrinsic range of line ratios is small, as even modest systematic uncertainties can significantly affect the inferred physical conditions.\medskip 

\hrule
     
 \vspace{-0.4cm}
 
\hrule

\newpage{}

\section{Synergies with current and future facilities}\vspace{-1em}
Deciphering the physics underlying the star-forming scaling relations connects to a refined view of the full baryon life cycle. In this context, we can particularly expect the following synergies: 
\vspace{0.2em}

\noindent \textbf{Atomic Gas: H\textsc{i} with SKAO/DSA/ngVLA:} 
The formation of stars involves the accretion, cooling, and conversion of gas from the circumgalactic medium through the atomic phase traced by H\textsc{i} into molecular gas traced by ALMA. SKAO will map this atomic reservoir and its connection to the circumgalactic medium over large areas and substantial look-back times, including the streams and plumes associated with gas accretion \citep{SKA}. Importantly, star formation may occur before a substantial, detectable molecular reservoir has formed, while CO itself may miss CO-dark H$_2$; molecular-line observations therefore provide only part of the star-formation cycle. Joint H\textsc{i} and molecular kinematics further constrain the rotation curve, and hence the gravitational potential, allowing the local shear, orbital time, and tidal field to be related directly to the conditions governing cloud collapse \citep{Meidt2018,Meidt2020,Kruijssen2019}. \vspace{0.2em} 

The ngVLA can also observe molecular lines, such as OH, H$_2$O masers, and even CO(1-0), at similar sensitivity and resolution as ALMA2040. The combination of the two facilities would therefore provide complementary constraints on the molecular ISM: ngVLA observations of low-frequency transitions and high-dipole-moment species would probe cold and dense gas, shocks, and maser activity, while ALMA2040 would provide the extensive suite of CO isotopologue and other molecular transitions needed to constrain temperature, density, and optical depth. \vspace{0.2em} 

\noindent\textbf{Stellar Populations and the ionized gas with ELT, and Roman:}
The molecular gas cannot be understood independently of the young stars that heat, ionize, and dynamically perturb it. The ELT, particularly through high-resolution integral-field spectroscopy with HARMONI, will provide spatially resolved measurements of the ionized ISM at tens-of-parsec and, for nearby galaxies, potentially even smaller physical scales \citep{ELT_HARMONI}. This will provide measurements of ionized-gas kinematics, electron density and temperature, metallicity, ionization structure, and the energetics of stellar feedback. Combining these observations with ALMA2040 molecular-gas maps will directly connect the properties of the young stellar population and ionized ISM to the temperature, density, and chemistry of the molecular gas, allowing environmental variations in the molecular gas to be linked to their stellar and feedback drivers. 
\vspace{0.2em}


Roman will provide a complementary view of the stellar component of nearby galaxies, combining wide-field near-infrared imaging with spectroscopy and the sensitivity required to resolve faint stellar populations. In particular, Roman will enable spatially resolved measurements of stellar mass, star-formation histories, and the structure and environment of nearby galaxies over substantially larger samples than currently possible \citep{Roman}. Proposed nearby-galaxy programmes envision mapping of hundreds of galaxies, extending to their outer disks and halos. 
\medskip 


\noindent \textbf{Interpreting Observations: Synergies with numerical simulations:}
Along with new observatories by the 2040s, we also expect advances in numerical galaxy simulations. Such simulations now enable more realistic CO chemistry and span sufficiently large regions to accurately describe the multiphase structure of the ISM in different environments \citep{Gong2020, Hu2022}. GMC-scale simulations are able to resolve the ISM down to regions close to the stellar surfaces, including all relevant physics and feedback processes (e.g., SILCC, \citealt{Walch2015}; STARFORGE, \citealt{Grudic2022}). Similarly, galactic-scale simulations now resolve individual molecular clouds either on high-resolution regions or over the entire galaxy, allowing us to study the dense ISM self-consistently in different galactic contexts (e.g.,\citealt{Thompson2024}; Rhea, \citealt{Goeller2025, Tress2026}). And cosmological zoom-in simulations build galaxies self-consistently with solar-mass resolutions with ever-increasing accuracy (eg. VESTRAL simulations). We can expect progress on all these fronts, merging the boundaries between GMC, galactic, and cosmological simulations, while steadily improving the physical prescriptions such as feedback and, importantly, chemical modelling of the dense ISM. Such improved models will even more strongly depend on accurate star-forming scaling relations across different scales and galaxy types and cross validation with observations. Moreover, new tools, including machine learning approaches (e.g. simulation-based inference),  will enable more robustly interpreting multi-line observations by implementing more physically motivated conditions (e.g. accounting for sub-beam density and temperature variations), constraining key molecular gas properties.  

\vspace{-0.6em}

\begin{tcolorbox}[boxsep=3pt,left=0pt,right=0pt,top=1pt,bottom=1pt,colframe=boxcolor,colback=boxcolor]

\textbf{Conclusion:} ALMA2040 will enable a qualitatively new experiment in galaxy evolution: measuring the physical state of molecular gas and the subsequent star-formation activity at matched spatial scales across a statistically representative galaxy population. Rather than testing whether molecular gas correlates with star formation, we will determine which physical properties of the gas predict star formation, how those properties depend on environment, and whether a universal law of star formation exists across the local galaxy population.
\end{tcolorbox}
\vspace{-2.2em}

\section*{Acknowledgements}\vspace{-1em}
This work was developed as part of the broader ALMA2040 community effort. We acknowledge the Lorentz Center and NOVA for supporting the community workshops and coordination activities that helped initiate and shape the ALMA2040 White Paper process.\\

\vspace{-1em}
\bibliographystyle{apj_mod}
{\bibliography{bibliography}}

\newpage{}
\noindent\textbf{Affiliations}

{
\footnotesize{
\noindent $^1$ Max-Planck-Institut for Astronomy, Königstuhl 17, 69171 Heidelberg, Germany, \\
$^2$ DARK, Niels Bohr Institute, University of Copenhagen, Jagtvej 155, 2200 Copenhagen, Denmark,\\
$^3$ ESO, Karl-Schwarzschild-Strasse 2, 85748 Garching, Germany,\\
$^{4}$ Leiden Observatory, Leiden University, PO Box 9513, NL-2300 RA Leiden, The Netherlands,\\
$^5$ Transdisciplinary Research Area (TRA) ‘Matter’ - Hertz Chair for Astrochemistry,  Universit\"{a}t Bonn, Am Propsthof 49, 53121 Bonn, Germany,\\
$^{6}$ INAF – Istituto di Radioastronomia, Via Piero Gobetti 101, 40129 Bologna, Italy,\\
$^{7}$ Centro de Astrobiología (CAB), CSIC-INTA, Ctra. de Ajalvir Km. 4, 28850, Torrejón de Ardoz, Madrid, Spain,\\
$^8$ Cardiff Hub for Astrophysics Research \&\ Technology, Department of Physics \&\ Astronomy, Cardiff University, Queens Buildings, Cardiff, CF24 3AA, UK,\\
$^9$ National Radio Astronomy Observatory, 800 Bradbury SE, Suite 235, Albuquerque, NM 87106 USA,\\
$^{10}$ Departamento de F\'{i}sica Te\'{o}rica y del Cosmos, Campus de Fuentenueva, Edificio Mecenas, Universidad de Granada, E-18071, Spain,\\
$^{11}$ Instituto Carlos I de F\'{i}sica Te\'{o}rica y Computacional, Facultad de Ciencias, E-18071, Granada, Spain,\\
$^{12}$ Universit\"{a}t Heidelberg, Zentrum f\"{u}r Astronomie, Institut f\"{u}r Theoretische Astrophysik, Albert-Ueberle-Str.\ 2, 69120 Heidelberg, Germany,\\
$^{13}$ Space Telescope Science Institute, 3700 San Martin Drive, Baltimore, MD 21218, USA,\\
$^{14}$ Universit\"{a}t Heidelberg, Interdisziplin\"{a}res Zentrum f\"{u}r Wissenschaftliches Rechnen, Im Neuenheimer Feld 225, 69120 Heidelberg, Germany, \\
$^{15}$ Centro de Astrobiología, CAB, CSIC-INTA, Spain,\\
$^{16}$ Department of Astronomy, University of Michigan, 1085 South University Avenue, Ann Arbor, MI 48109, USA,\\
$^{17}$ Instituto de F\'isica Fundamental, CSIC, Calle Serrano 123, 28006 Madrid, Spain,\\
$^{18}$ Observatorio Astronómico Nacional (IGN), C/Alfonso XII 3, Madrid E-28014, Spain,\\
$^{19}$ INAF, Arcetri Astrophysical Observatory, Largo Enrico Fermi 5, I-50125 Florence, Italy,\\
$^{20}$ INAF - Osservatorio Astronomico di Trieste, Via G. B. Tiepolo 11, I-34143 Trieste, Italy,\\
$^{21}$ IFPU - Institute for Fundamental Physics of the Universe, via Beirut 2, I-34151 Trieste, Italy,\\
$^{22}$ Department of Physics and Astronomy, University of Kentucky, 506 Library Drive, Lexington, KY 40506, USA,\\
$^{23}$ Department of Astrophysics, University of Vienna, Türkenschanzstraße 17, 1180 Vienna, Austria,\\
$^{24}$ UK ALMA Regional Centre Node, Jodrell Bank Centre for Astrophysics, Department of Physics and Astronomy, The University of Manchester, Oxford Road, Manchester M13 9PL, UK\\
}
}
\vspace{-1em}
\end{document}